\documentclass[aps,prl,twocolumn,showpacs,preprintnumbers,amsmath,amssymb,superscriptaddress]{revtex4}

\usepackage{graphicx}
\usepackage{dcolumn}

\newcommand{\be}{\begin{equation}}
\newcommand{\ee}{\end{equation}}
\newcommand{\bea}{\begin{eqnarray}}
\newcommand{\eea}{\end{eqnarray}}
\newcommand{\bm}[1]{\mathbf{#1}}

\newcommand{\lp}{\left(}
\newcommand{\rp}{\right)}

\newcommand{\ty}[1]{\mbox{\tiny #1}}

\begin{document}

\noindent
{\textbf{Comment on "Electron screening and excitonic condensation in double-layer graphene systems"} \\[0.2 cm]}
\ \
\begin{abstract}

\end{abstract}

Recently we\cite{us,yogesh} predicted that a clean\cite{usdisorder} double-layer graphene system with a small layer
separation and nested electron and hole Fermi spheres induced by external gates
will exhibit pair-condensation\cite{yogesh} at temperatures $\sim 0.1 \epsilon_{F}$\cite{us}, where $\epsilon_{F}$
is the Fermi energy of the gate-induced electron gases.  A very different conclusion was
reached in a recent preprint\cite{them} by M. Y. Kharitonov and K. Efetov who conclude that the maximum pair-condensation
temperature is $\sim 10^{-7} \epsilon_{F}$.  The stark contrast between these estimates is a combined consequence of
the different approximations used for the effective interlayer interaction
and the extreme sensitivity of the condensation temperature to the strength of this interaction when it is weak.

Kharitonov and Efetov's estimate for the critical temperature
is based on using a mean-field-theory linearized $T_c$ equation
combined with Thomas-Fermi screened interlayer interactions.
As already discussed in our original paper, we agree with Kharitonov and Efetov
that low transition temperature estimates follow from this procedure:
{\em `` Screening and other beyond-mean-field induced-interaction effects
are difficult to describe. In the case of
weakly interacting atomic gases induced interaction effects can[18] either increase or decrease $T_c$, depending on
the number of fermion flavors g. For the present Coulomb
interaction case, a static Thomas-Fermi screening approximation with normal state screening wavevectors
reduces interaction strengths very substantially when spin
and valley degeneracies (g = 4) are included. $\cdot \cdot \cdot$ On the other hand, when the screening
wavevectors are evaluated in the condensed state there
is little influence on $T_{KT}$ at small $k_{\ty F} d$ both because the
large gap weakens screening and because $T_{KT}$ is proportional to the Fermi energy and not to the interaction
strength in this limit. All this leads us to suspect that at
low-temperatures there is a first-order phase transition
as a function of layer separation $d$ between condensed
and electron-hole plasma states, similar to the transitions
studied experimentally[19] in quantum Hall exciton condensates
and theoretically[20] in parabolic band bilayers.''}
References [18-20] in Ref.\cite{us} are listed here as References [5-7].
We elaborate on our views below.

Corrections to mean-field-theory estimates of $T_c$ can be treated rigorously only in the limit of
weak short-range interactions\cite{heiselberg}.  For the circumstance\cite{caveat} treated in Ref.[~\onlinecite{heiselberg}],
induced interactions corrections reduce $T_c$ for $g=2$, but enhance $T_c$ for $g=4$.
When interlayer interactions are strong, the static screening approximation used by Kharitonov and Efetov does not represent
a systematic improvement on mean-field theory even if the number of fermion flavors is large.
Nevertheless, our assessment of the importance of screening for spontaneous bilayer coherence follows
from the observation that two distinct solutions emerge when mean-field theory
is consistently combined with Thomas-Fermi screening theory, the low $T_c$ solution discussed by
Kharitonov and Efetov and a high $T_c$ solution.  In our opinion the low $T_c$ solution
is unlikely to be physically realistic at any layer separation; we predict instead that there is a first-order
phase transition as a function of layer separation between a state without spontaneous
coherence and a high-$T_c$ state as discussed in Ref.[~\onlinecite{us}].

The influence of screening on the interlayer and intralayer interactions in bilayer systems
was considered long ago in the context of semiconductor based bilayers. In the absence of interlayer coherence it is common practice to
estimate the screened potentials using the random phase approximation. However, in the condensed state it is more difficult
to consistently account for dynamical screening and we resort to the Thomas Fermi approximation.
Within that approximation a straightforward calculation for the screened potentials of a balanced bilayer yields
\begin{widetext}
\begin{equation}
V^{sc} =
\frac{1}{2} \frac{V_{S} + V_{D}}{1 + \Pi_+ (V_{S}+V_{D})}
\pm \frac{1}{2} \frac{V_{S} - V_{D}}{1 + \Pi_- (V_{S}-V_{D})}
\end{equation}
\end{widetext}
where the plus sign corresponds to the screened intralayer potential and the minus sign to the
screened interlayer potential.
Here $V_{S}=2\pi e^2/ \varepsilon q$ is the 2D Coulomb interaction between electrons in the
same layer, $V_{D}=\exp(-qd) V_{S}$ is the interlayer interaction with $d$ being the distance between layers, and
\begin{equation}
\label{eq:pi}
\Pi_{\pm} = \Pi_{S} \pm \Pi_{D}.
\end{equation}
In Eq.(~\ref{eq:pi}),
$\Pi_{S}$ and $\Pi_{D}$ are respectively the static intralayer and interlayer polarization operators that relate a shift in the
chemical potential (relative to the electrostatic potential) in one layer to the corresponding density in the same-layer ($S$) or in the
opposite layer ($D$=different).  We note that $\Pi_D$ differs from zero only in the condensed state. Only then does the density in one layer depend
on the chemical potential of the other layer.

In the normal state
$\Pi_{D}$ vanishes and $\Pi_{S} = g \nu_0$ where $\nu_0$ is the density-of-states at the Fermi energy per valley and spin in either layer and $g=4$.
It then follows that the screened interlayer interaction is
\begin{equation}
V_{D}^{sc} = \frac{1}{\nu_0 g} \frac{q_{\ty{TF}} \exp(-qd)}{q + 2 q_{\ty{TF}}  + q_{\ty{TF}}^2 \left[1-\exp(-2qd)\right]/q } \le \frac{1}{2 \nu_0 g}
\end{equation}
where $q_{\ty{TF}} = 2\pi e^2 g \nu_0/\varepsilon $ is the Thomas-Fermi wave vector.
We have calculated the mean field critical temperature with this interaction and found that
$T_c \sim 6\cdot 10^{-5} \epsilon_{F}$.
Although not in complete agreement with the estimate of Kharitonov and Efetov, we
fully agree that this analysis predicts a small value of $T_c$ rendering the pairing phenomena
improbable in realistic disordered bilayer systems\cite{usdisorder}.

Quite a different conclusion is reached if
Thomas-Fermi screening theory is applied consistently in the condensed state.
While $\Pi_-$ retains it normal state value $g\nu_0$,
\be
\Pi_{+} = g \sum_\bm{k} \partial_{E_k} n_{\ty{F}}\lp E_k \rp         \label{eq:pi_plus}
\ee
is considerably reduced. Here $E_k$ is the energy
of a quasiparticle in the condensed state. In fact, due to the excitation gap $\Pi_+$ vanishes at zero temperature.
Surprisingly at low temperatures $V^{sc}_{D} \approx (V_{S} + V_{D})/2$ and hence,
the mean-field theory gaps tend to be even
larger than when screening is neglected.  Although we do not take this argument for
enhanced interactions particularly seriously, we conclude that interlayer
phase coherence is completely consistent with naive Thomas Fermi screening considerations.

At high temperatures $\Pi_+ \sim g\nu_0$ as well, screening substantially diminishes the order parameter
and the system undergoes a first order phase transition into the normal phase.
It follows from Eq.(\ref{eq:pi_plus}) that the transition to the normal state occurs when $\Delta \sim T$.
As a consequence, the mean field $T_c$ overestimates the
real critical temperature. Nevertheless,
mean field theory is adequate for determining the Kosterlitz-Thouless temperature, $T_{\ty{KT}}$,
as long as $\Delta(T_{\ty{KT}}) \gtrsim T_{\ty{KT}}$. We expect this condition to be satisfied
in the strongly interacting regime where $\Delta(T_{\ty{KT}}) \sim \Delta(0)$.

In the absence of experimental studies of graphene double-layer systems,
we can attempt to take guidance from two other partially related systems in
which electron-pair condensation has been achieved.
A combination of experiment, quantum Monte-Carlo,
and various approximate calculations suggests that
cold-atom fermions with strongly attractive interactions
are reasonably well described at low-temperatures by including
collective fluctuations around the BCS mean-field theory state.
Moreover, at unitarity numerical calculations find that $T_c \approx 0.2 E_{\ty{F}}$ \cite{unitarityMonteCarlo}.

A closer comparison is possible between
graphene-bilayers and semi-conductor based bilayers
in the quantum Hall regime.  In this case the Kharitonov and Efetov estimate would
yield $T_c \to 0$ because the Thomas-Fermi screening wavevector diverges in concert with the
zero-width of 2D Landau bands.  Although experiments are not yet able to
extract unambiguous signals of a finite-T phase transition, it is nevertheless clear that
spontaneous coherence anomalies begin to emerge below the Kosterlitz-Thouless temperature
estimated by a $T=0$ mean-field-theory phase stiffness calculation.
Moreover it is known\cite{yangDassarmaMacdonald} that spontaneous interlayer phase
coherence occurs at small layer separations irrespective of the number of fermion flavors.

Admittedly, an accurate estimate for the critical temperature in the strongly interacting
regime $k_{\ty{F}} d \lesssim 1$ does not exist at present\cite{monteCarlo}. Nevertheless
we feel that the attractive features of graphene bilayers: high carrier densities, high electron and hole Fermi energies, and nearly-perfect particle-hole symmetry
suggest that interesting and novel phenomena are likely in this system.
We hope that the intriguing thermodynamic and transport aspects of bilayer graphene
as well as its potential use in applications encourage experimentalists to construct this system
and explore its electrical properties in different geometries\cite{jungjung} .     \\[0.2 cm]

\begin{tabular}{ll}
& R. Bistritzer, H. Min, J. J. Su, and A.H. MacDonald \\
& Department of Physics, \\
& The University of Texas at Austin, \\
& Austin Texas 78712\\
\end{tabular}


\begin{thebibliography}{999}
\bibitem{caveat}  The model studied in this paper has attractive interactions between
all particles, whereas a bilayer has attractive interlayer electron-hole interactions
and repulsive intralayer electron-electron interactions.
\end{thebibliography}
\end{document}